# Circular Polarizer Realized by a Single layer of Planar Chiral Metallic Nanostructure

Weimin Ye[1]*, Xiaodong Yuan[1], Chucai Guo[1], Jianfa Zhang[1], Biao Yang[1], Ken Liu[1], Zhihong Zhu[1] & Chun Zeng[1]

[1] College of Optoelectronic Science and Engineering, National University of Defense Technology, Changsha, Hunan 410073, P.R. China.

*Correspondence to: wmye72@126.com

**Abstract**: As a basic optical element, circular polarizer plays significant roles in signal transmission, measurements and life science microscopy. Three-dimensional (3D) chiral structures have been thought to be necessary to realize circular polarizers. Here we demonstrate theoretically and experimentally for the first time that a high-efficiency circular polarizer could be realized by a single layer of planar 2D chiral structure. Our proposed circular polarizer is based on unidirectional polarization conversion instead of circular polarization stop bands. Since two-dimensional planar structures present obvious advantage for fabrication and integration on chip, the proposed circular polarizer is of great interest in integrated optics and microscopy. It provides a novel scheme to manipulate polarizations of light wave, as well as Terahertz wave and microwave.

The fabrication of circular polarizer has been challenging. 3D spiral (*1*), bi-chiral (*2*) and gyroid (*3*) photonic crystals have been demonstrated to have circular polarization stop bands. A miniature chiral polarizing beamsplitter was realized by gyroid photonic crystals (*4*). Metamaterials have also been employed to realize circular polarizers. Broadband circular polarizers in the optical regime have been demonstrated by using gold helix photonic metamaterial (*5*) and twisted optical metamaterials (*6*). Up to our knowledge, all reported circular polarizers were designed that the incident light and transmitted light possess identical polarization characteristics. Consequently, eigenmodes of the circular polarizers should be circular polarization. Under this requirement circular polarizers have to be 3D chiral structures.

In this manuscript we propose a novel scheme that does not need 3D structure to realize circular polarizers. In this scheme incident right-circular polarization (RCP) becomes left-circular polarization (LCP) when transmitted, while incident LCP is reflected and keeps its polarization as LCP. Using Lorentz reciprocity theorem (*7*), we know that incident LCP from the opposite direction will become RCP when transmitted if the circular polarizer is made of reciprocal media. So, this new kind of circular polarizer is based on unidirectional polarization conversion (UPC) rather than circular polarization stop bands. A similar linear polarizer has been realized by cascading two nonparallel gratings made of isotropic and linear media (*8*). By a straightforward discussion based on the scattering matrix method, we know that the circular polarizer could be realized by a single layer of symmetrical planar chiral structure which supports four linearly polarized eigenmodes (See Supplementary Materials). It is worth noticing that planar chiral metallic nanostructures (*9-11*) have been designed to obtain asymmetric transmission of circular polarizations (CPs). In those structures asymmetric transmittance is described quantitatively by the asymmetric polarization conversion. But, circular polarization conversion is weak because those planar chiral nanostructures are thin. Anisotropy of loss (*11*) is thought to be crucial for achieving the asymmetric transmission. Therefore, difference between transmittances of RCP and LCP is small.

Here we focus on the optical communication band and propose a plasmonic circular polarizer based on a single layer of two-dimensional (2D) chiral structure with mirror symmetry about the horizontal(x-y) plane. As shown in Fig. 1A, the circular polarizer consists of a periodic plane array of asymmetrical L-shaped gold particles in a square lattice with period $a$. All L-shaped particles are buried in silica ($SiO_2$). The two arms of L-shaped particles, oriented in the X- and Y- directions, have identical length $L$ and thickness $H$, different widths $W_x$ and $W_y$ as shown in Fig. 1B. When the thickness $H$ approaches infinity, the asymmetrical L-shaped gold periodic array works as waveguides and supports four kinds of linearly polarized eigenmodes. Among them there are 45°-like and -45°-like linear polarizations which are weakly dispersive plasmonic modes. They are induced mainly by surface charges located at the two end facets. The other two dispersive waveguide modes are Y-like and X-like linear polarizations induced mainly by surface charges located along X-directional arm and Y-directional arm respectively. The latter two modes are similar to the eigenmodes in metallic waveguide and have different cut-off frequencies inversely proportional to the widths of waveguide. The key point of the planar circular polarizer is the Four-linearly polarized Mode Interference (FMI). So, the thickness of the planar circular polarizer is much higher than that of the reported planar chiral metallic nanostructures (See Supplementary Information).

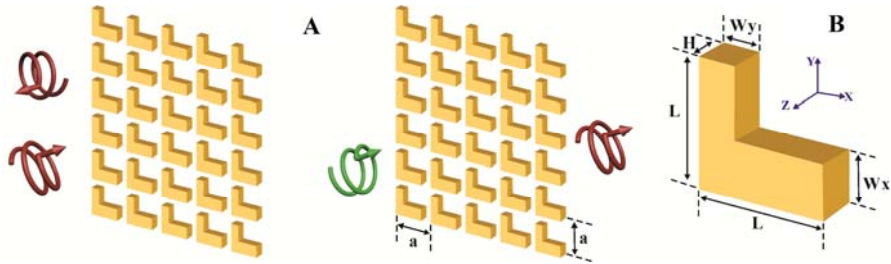

**Fig. 1**. Proposed circular polarizer based on UPC. (**A**) The circular polarizer consists of a periodic plane array of asymmetrical L-shaped gold particles in a square lattice with period a. Where, an incident right-circular polarization (RCP) becomes a left-circular polarization (LCP) when transmitted, an incident LCP is reflected and keeps its polarization as LCP. (**B**) The structure of asymmetrical L-shaped gold particles with identical length $L$ and thickness $H$, different widths $W_x$ and $W_y$.

Figure 2 shows the transmittance and reflectance spectra of RCP and LCP at normal incidence to the proposed circular polarizer calculated by COMSOL. Geometrical parameters of L-shaped gold particles buried in $SiO_2$ are $a$=730nm, $L$=580nm, $W_x$=246nm, $W_y$=160nm, $H$=360nm. It is worth noticing in the relative polarization transmittance spectra shown in Fig2B that for a normally incident light with wavelength within the interval [1460nm, 1600nm], the circular polarizer only supports high transmission rate of RCP incidence to LCP transmittance. While, transmissions of RCP to RCP, LCP to RCP and LCP to LCP are inhibited. So, Fig.2A shows that within the wavelength interval, the transmittance of LCP is less than 0.015, and that of RCP is greater than 0.8. The transmitted light of an incident RCP is nearly LCP, whose ellipticity is less than -0.81. Moreover, reflectance spectra (Fig. 2C) and relative polarization reflectance spectra (Fig. 2D) show that the reflectance of LCP with wavelength within the same interval is greater than 0.9. The reflected light is nearly LCP, whose ellipticity is less than 0.8. Due to the optical loss in gold, the transmittance of the circular polarizer is less than 0.95. The absorbance of the circular polarizer in the optical communication band keeps less than 10%. Furthermore, by solving eigenmodes of asymmetrical L-shaped gold periodic array waveguides, which have the same geometrical parameters in the horizontal plane as the circular polarizer, we know that both Y-like and X-like linearly polarized eigenmodes are cut off for a light with the wavelength approaching or greater than 1800nm. New guided modes appear for a light with the wavelength approaching or less than 1200nm (See Supplementary Materials). In these two cases, the Four-linearly polarized Mode Interference in the circular polarizer does not work effectively. Therefore, the proposed circular polarizer can not operate in these cases.

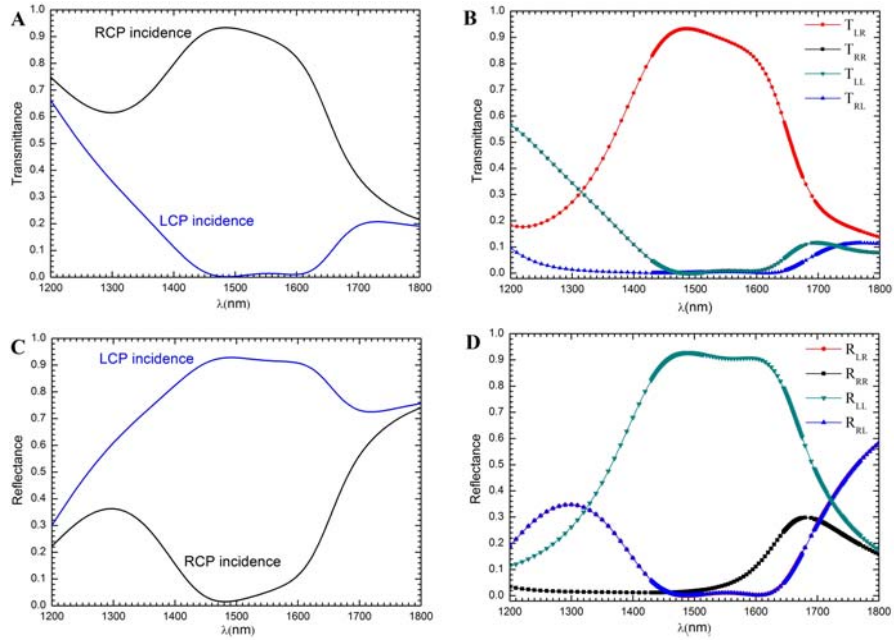

**Fig. 2**. Calculated spectra of the designed circular polarizer. (**A**) Transmittance and (**B**) relative polarization transmittance spectra of RCP and LCP at normal incidence to the proposed circular polarizer respectively. Where, $T_{LR}$ ($T_{LL}$) and $T_{RR}$ ($T_{RL}$) denote the LC- and RC-polarized transmittance of an incident RCP (LCP). (**C**) Reflectance and (**D**) relative polarization reflectance spectra of them. Where, $R_{LR}$ ($R_{LL}$) and $R_{RR}$ ($R_{RL}$) denote the LC- and RC-polarized reflectance of an incident RCP (LCP). It is obvious that the circular polarizer converts efficiently an incident RCP into a transmitted LCP and reflects an incident LCP into a LCP.

To demonstrate the feasibility of the proposed circular polarizer, a sample was fabricated on a 0.5mm thick double-side polished $SiO_2$ substrate. A 5nm thick Chromium (Cr) adhesion layer was first deposited on the substrate using e-beam evaporator. A 360nm thick gold layer was sequentially deposited using DC sputter. Then, the focused ion beam (FIB) system (FEI Co. Helios Nanolab 600i. 30 keV Ga ions) was used to mill the gold layer to obtain the designed periodic array of asymmetrical L-shaped gold particles. To mimic L-shaped gold particles buried in $SiO_2$, silica refractive index matching liquid (Carigille Lab Series AA-1.4580) was added on the patterned gold layer. Finally, another $SiO_2$ wafer same as the substrate was covered on the sample for the purpose of protection. The footprint of the sample was about 43μm by 43μm. Fig. 3 shows SEM images of the fabricated asymmetrical L-shaped gold particles. The sample shown in Fig 3A and 3B was tilted by 52 degree and 0 degree respectively. In order to guarantee the gold layer to be milled completely, the SiO$_2$ substrate was also milled a little bit. It was obvious in Fig3A that the cross section of fabricated L-shaped gold particle is trapezoid. As we know, a periodic array of L-shaped gold particles with a trapezoid cross section could not work efficiently as waveguides. The trapezoid cross section is harmful for Four-linearly polarized Mode Interference in the structure. As a result, the performance of the fabricated sample will be deteriorated, especially for the working bandwidth (See Supplementary Materials). Therefore, our experiment can only serve the purpose of proof-of-concept.

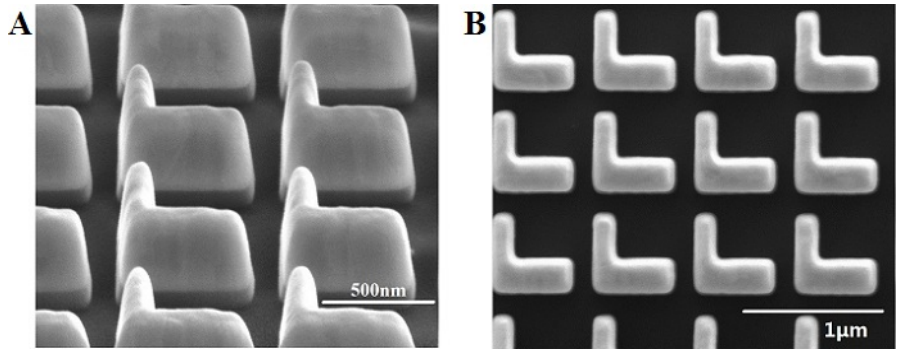

**Fig. 3**. SEM images of the asymmetrical L-shaped gold particles milled by FIB. The sample were tilted by 52 degree (**A**) and 0 degree (**B**), respectively. In order to guarantee the gold layer was milled completely, the SiO$_2$ substrate was also milled. The cross section of fabricated L-shaped gold particle was trapezoid.

We tested the sample with a broadband light source (Spectral Products ASBN-W050F). The output light from it was focused by a lens and sequentially passed through an aperture, a linear polarizer (Thorlabs GTH5M), a achromatic quarter-wave plate (Thorlabs AQWP10M-1600), an object lense (Olympus UPLFLN 10X2 N.A. 0.3), the sample, a second object lense, a second achromatic quarter-wave plate, a second linear polarizer, and finally coupled into a multimode fiber connected with an optical spectrum analyzer (Yokogawa AQ6370C). The fast axes of the two quarter-wave plates were set perpendicular to each other. The polarization directions of two linear polarizers were also set perpendicular to each other. Changing the directions of two fast axes of quarter-wave plates with fixed polarization directions of linear polarizers, we could measure the relative polarization transmittance spectra of RCP and LCP at normal incidence to the sample. In Fig.4A, the solid lines show the measured transmittance spectra of RCP and LCP at normal incidence to the sample. And the dashed lines are the calculated transmittance spectra of the circular polarizer consisting of L-shaped gold particles with a trapezoid cross section (more details in Supplementary Material). The experimental and theoretical results are basically consistent. While, comparing the measured relative polarization transmittance spectra in Fig. 4B with the corresponding calculated spectra, we can find that the measured transmittance of RCP to LCP is relatively high as expected, but the transmittances of RCP to RCP and LCP to RCP are not suppressed efficiently due to manufacturing imperfections (geometric deviations and inhomogeneity) of the L-shaped gold particles. So, the measured working bandwidth and transmission contrast ratio between transmittances of RCP and LCP are smaller than the theoretical results.

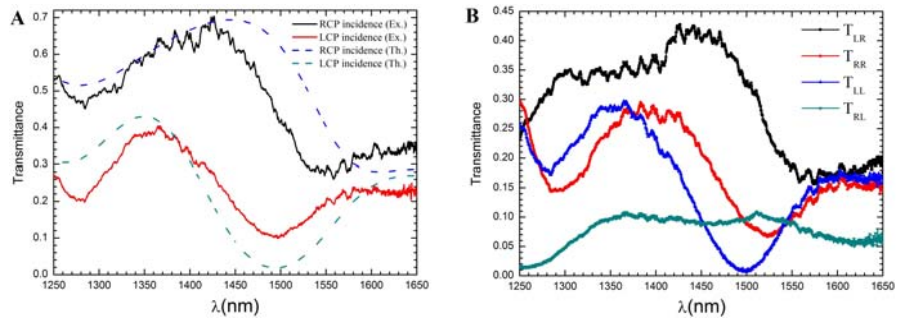

**Fig. 4**. Measured spectra of the fabricated circular polarizer. Measured transmittance (**A**) and relative polarization transmittance (**B**) spectra of RCP and LCP at normal incidence to the sample. The dashed lines in a are the calculated transmittance of a circular polarizer consisting of L-shaped gold particles with a trapezoid cross section (See Supplementary Materials).

In conclusion, we demonstrated that a high-efficiency circular polarizer could be realized by a single layer of symmetrical planar 2D chiral structure by theory and experiments. Different from reported circular polarizers, the proposed circular polarizer is based on unidirectional polarization conversion. That is, RCP (LCP) can only pass through it in one (the opposite) direction and becomes LCP (RCP) when transmitted. The reflected CP keeps its circular polarization. So, for a linearly polarized or unpolarized light incident to the circular polarizer, the transmitted and reflected light will be identical circular polarizations. It provides a novel scheme to manipulate polarizations of light.

**Acknowledgments:** This work was funded by the National Natural Science Foundation of China under Grant 11374367. We acknowledge the Center of Material Science of National University of Defense Technology for help in FIB Fabrication. We also thank Wei Xu for help in COMSOL simulation, Wei Chen for help in fabrication. All other authors declare no competing financial interests.

**Supplementary Materials:**

Materials and Methods

**Supplementary Materials:**

**Materials and Methods:**

1. Scattering Matrix of the Proposed Perfect Circular Polarizer

Figure S1 shows a general optical circuit of a circular polarizer. Polarization modes R± and L± in the figure denote the electric fields of right-circular polarization (RCP) and left-circular polarization (LCP) and can be written as

$$\boldsymbol{R}^{\pm} = (\boldsymbol{e}_x \pm i\boldsymbol{e}_y)e^{i(\pm\beta z - \omega t)}, \qquad \boldsymbol{L}^{\pm} = (\boldsymbol{e}_x \mp i\boldsymbol{e}_y)e^{i(\pm\beta z - \omega t)}. \qquad (1)$$

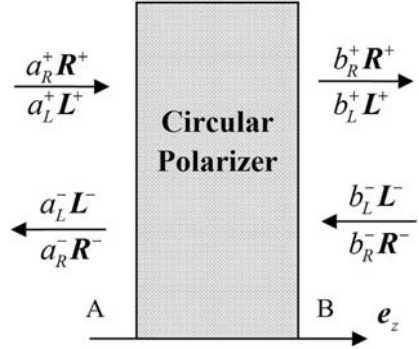

Fig. S1 General optical circuit of circular polarizer.

The relationship between corresponding amplitude coefficients of the polarization modes can be described by scattering matrix as

$$\begin{bmatrix} b_R^+ \\ b_L^+ \\ a_L^- \\ a_R^- \end{bmatrix} = \boldsymbol{S} \begin{bmatrix} a_R^+ \\ a_L^+ \\ b_L^- \\ b_R^- \end{bmatrix} = \begin{bmatrix} S_{R^+R^+} & S_{R^+L^+} & S_{R^+L^-} & S_{R^+R^-} \\ S_{L^+R^+} & S_{L^+L^+} & S_{L^+L^-} & S_{L^+R^-} \\ S_{L^-R^+} & S_{L^-L^+} & S_{L^-L^-} & S_{L^-R^-} \\ S_{R^-R^+} & S_{R^-L^+} & S_{R^-L^-} & S_{R^-R^-} \end{bmatrix} \begin{bmatrix} a_R^+ \\ a_L^+ \\ b_L^- \\ b_R^- \end{bmatrix}. \qquad (2)$$

By applying Lorentz reciprocity theorem, we conclude that

$$S_{R^+R^+} = S_{R^-R^-}; \quad S_{L^+L^+} = S_{L^-L^-}; \quad S_{L^+R^+} = S_{R^-L^-}; \quad S_{R^+L^+} = S_{L^-R^-}. \qquad (3)$$

Especially, for a symmetrical structure with mirror symmetry about the horizontal(x-y) plane, the matrix elements have to satisfy the following equations

$$S_{R^+R^+} = S_{L^-L^-}; \quad S_{L^+L^+} = S_{R^-R^-}; \quad S_{R^-R^+} = S_{L^+L^-}; \quad S_{L^-R^+} = S_{R^+L^-}; \quad S_{R^-L^+} = S_{L^+R^-}; \quad S_{L^-L^+} = S_{R^+R^-}. \qquad (4)$$

Therefore, scattering matrix of a symmetrical and reciprocal planar structure is block symmetrical matrix.

For a symmetrical and reciprocal perfect circular polarizer based on unidirectional polarization conversion (UPC), its scattering matrix can be written as

$$S = \begin{bmatrix} 0 & 0 & 0 & e^{i\delta} \\ e^{i\alpha} & 0 & 0 & 0 \\ 0 & e^{i\delta} & 0 & 0 \\ 0 & 0 & e^{i\alpha} & 0 \end{bmatrix}. \tag{5}$$

Where, α is the transmission phase delay of an incident RCP which becomes LCP when transmitted. And δ is the reflection phase delay of an incident LCP which keeps its polarization as LCP when reflected. This matrix has four eigenvectors,

$$V_1 = \begin{bmatrix} 1 \\ e^{i\varphi} \\ 1 \\ e^{i\varphi} \end{bmatrix}; \quad V_2 = \begin{bmatrix} 1 \\ -e^{i\varphi} \\ 1 \\ -e^{i\varphi} \end{bmatrix}; \quad V_3 = \begin{bmatrix} 1 \\ ie^{i\varphi} \\ -1 \\ -ie^{i\varphi} \end{bmatrix}; \quad V_4 = \begin{bmatrix} 1 \\ -ie^{i\varphi} \\ -1 \\ ie^{i\varphi} \end{bmatrix}; \quad \varphi \triangleq \frac{\alpha - \delta}{2} \tag{6}$$

There are four eigenmodes corresponding to the four eigenvectors. At the side B in Fig.S1, eigen-electric fields of the four modes propagating along z direction are

$$\boldsymbol{E}_1 = R_B^+ + e^{i\varphi} L_B^+ = 2e^{i\varphi/2} e^{i(\beta z - \omega t)} \boldsymbol{e}_{\frac{\varphi}{2}}; \qquad \boldsymbol{E}_2 = R_B^+ - e^{i\varphi} L_B^+ = 2ie^{i\varphi/2} e^{i(\beta z - \omega t)} \boldsymbol{e}_{\frac{\varphi}{2}+\frac{\pi}{2}}; \tag{7}$$

$$\boldsymbol{E}_3 = R_B^+ + ie^{i\varphi} L_B^+ = 2e^{i\pi/4} e^{i\varphi/2} e^{i(\beta z - \omega t)} \boldsymbol{e}_{\frac{\varphi}{2}+\frac{\pi}{4}}; \quad \boldsymbol{E}_4 = R_B^+ - ie^{i\varphi} L_B^+ = 2e^{-i\pi/4} e^{i\varphi/2} e^{i(\beta z - \omega t)} \boldsymbol{e}_{\frac{\varphi}{2}-\frac{\pi}{4}}; \tag{8}$$

$$\boldsymbol{e}_{\frac{\varphi}{2}} \triangleq \cos\frac{\varphi}{2}\boldsymbol{e}_x + \sin\frac{\varphi}{2}\boldsymbol{e}_y. \tag{9}$$

It is obvious that these are four linear polarizations with adjacent polarization directions separated by 45 degrees. So, a circular polarizer based on UPC could be realized by a single layer of planar chiral structure which supports four linearly polarized eigenmodes when its thickness approaches infinite. And the four-mode interference is crucial for achieving the circularly polarized light based on UPC.

2. Eignmodes of Asymmetrical L-shaped Gold Periodic Array Waveguides

When the thickness of the gold particles approaches infinity, the asymmetrical L-shaped gold periodic array becomes a periodic array of waveguides. We use COMSOL Multiphysics 2D finite-element-based electromagnetic solver to calculate eignmodes of the waveguide array. Geometrical parameters of the L-shaped gold are a=730nm, L=580nm, Wx=246nm, Wy=160nm, respectively. The refractive index of $SiO_2$ is 1.46. The relative permittivity of gold is given by Drude model. That is

$$\varepsilon_r(\text{Au}) = 1 - \frac{\omega_p^2}{\omega(\omega + i\omega_\tau)}, \quad \omega_p = 1.37 \times 10^{16}\,\text{rad/s}, \quad \omega_\tau = 4.08 \times 10^{13}\,\text{rad/s} \tag{10}$$

Figure.S2a and S2b show the real part and imaginary part of the effective mode indexes of the four modes supported by an asymmetrical L-shaped gold periodic array waveguides with the lowest cut-off frequencies. All of the four modes are linear polarizations.

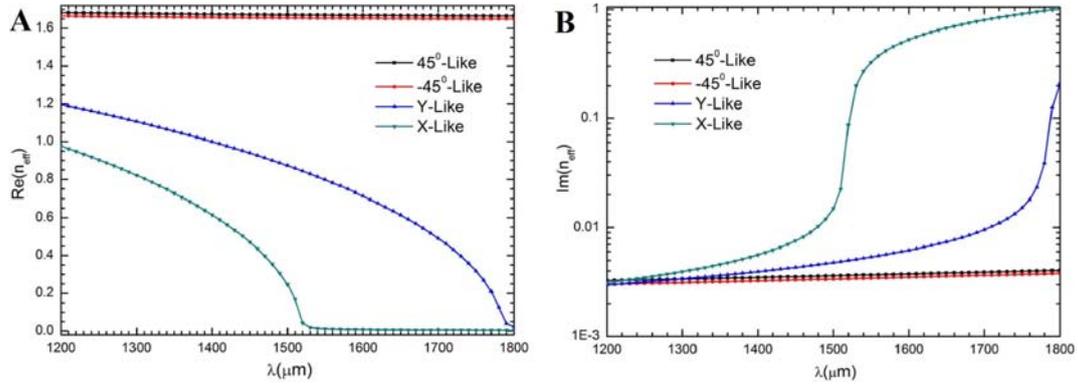

Fig.S2 Effective mode index Spectra of Four Eignmodes. Real part (**A**) and Imaginary part (**B**) of effective mode index $n_{eff}$ of the four eignmodes supported by asymmetrical L-shaped gold periodic array waveguides with lowest cut-off frequencies. The mode names are based on their transverse electric filed vector distribution shown in Fig.S3. The geometrical parameters of L are a=730nm, L=580nm, Wx=246nm, Wy=160nm, respectively.

The names of them are based on their transverse electric filed vector distributions at wavelength of 1500nm shown in Fig.S3. Among them two weakly dispersive plasmonic modes are named as 45°-like and -45°-like modes because they are induced mainly by surface charges located at the two end facets. The other two dispersive waveguide modes are named as Y-like and X-like modes because they are induced mainly by surface charges located along X-directional and Y-directional arms respectively. The latter two modes are similar to eigenmodes in metallic waveguide and have different cut-off frequencies inversely proportional to the widths of waveguide. Cut-off wavelengths of X-like and Y-like linearly polarized eigenmodes are about 1520nm and 1790nm, respectively. There are new guided modes (not shown) for a light with the wavelength smaller than 1200nm.

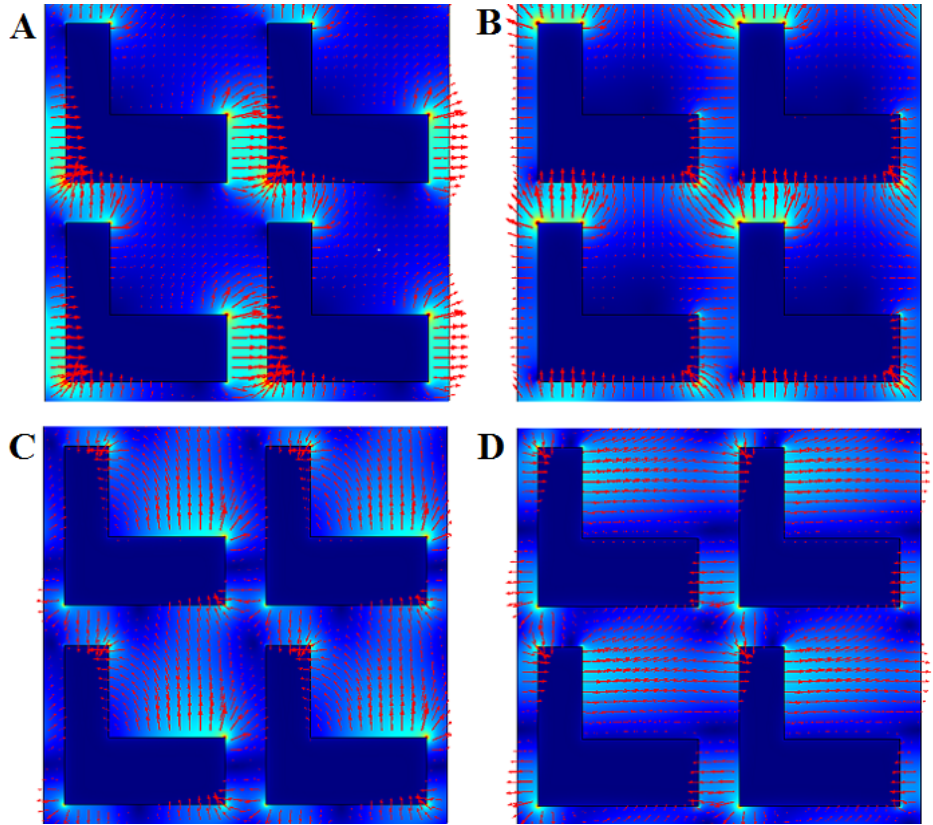

Fig.S3 Transverse electric filed vector distribution of Four Eignmodes. (**A**)-(**D**) Transverse electric filed vector distribution of four waveguide modes named 45°-like (**A**), -45°-like (**B**), Y-like (**C**) and X-like (**D**) linear polarizations for a wavelength of 1500nm. Their effective mode indexes are equal to 1.6714, 1.6553, 0.8736 and 0.2486 as in Fig.S2a respectively.

3. Scattering Matrix Model of the Proposed Circular Polarizer

By only considering the four eigenmodes (shown in Fig.S3) in asymmetrical L-shaped gold periodic array waveguides, we can use a model of scattering matrix shown in **Fig.S4** to analyze the proposed circular polarizer. Where, scattering matrix of the interface between semi-infinite $SiO_2$ and asymmetrical L-shaped gold periodic array waveguides could be calculated by using the port boundary in COMSOL finite-element-based electromagnetic solver.

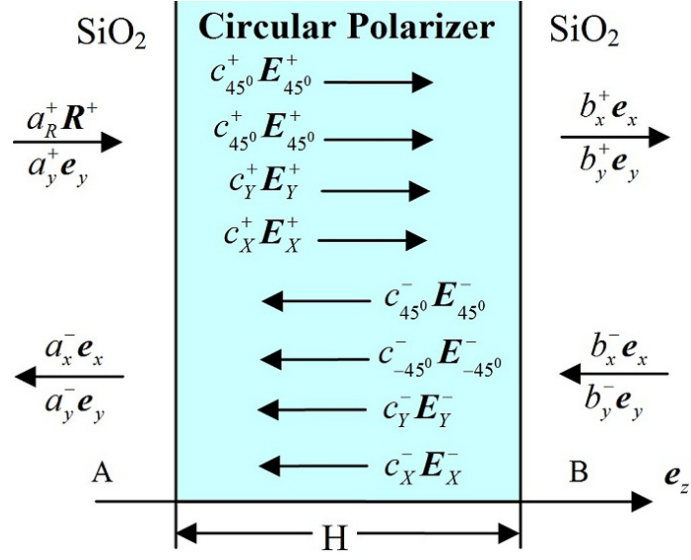

Fig.S4 Scattering matrix model of the proposed circular polarizer. The electric fields in the circular polarizer are composed only of four eigenmodes of the respective gold periodic array waveguides shown in Fig.S3.

For the normally incident RCP and LCP at wavelength 1500nm, Fig.S5a and S5c show the transmittance and reflectance of RCP and LCP through the proposed circular polarizer v.s. thicknesses $H$ respectively. Since it is the electric field vector summation contributes to the far field polarization, the ellipticities of the transmitted and reflected light are defined by their averaged electric fields in a periodic cell. In our calculation the ellipticity of polarized light with transverse electric filed proportional to $e_x \pm i e_y$ is supposed to be $\pm 1$. Fig.S5b shows the ellipticity of transmitted light for an incident RCP v.s. $H$. And, Fig.S5d shows that of reflected light for an incident LCP. It is not hard to find in the four figures that the results calculated by using scattering matrix (denoted by lines) agree well with those by using COMSOL (denoted by symbols). Based on these results, we can draw three useful conclusions. The first is that the proposed single layer of symmetrical planar chiral structure with a thickness equal to 360nm is an efficient circular polarizer for a normally incident light at wavelength 1500nm. The second is that the planar circular polarizer is realized by employing unidirectional polarization conversion. The last is that the mechanism behind the planar circular polarizer is the Four-linearly polarized Mode Interference (FMI). So, to realize the planar circular polarizer we need much higher thickness of L-shaped gold particle than the reported planar chiral metallic nanostructures.

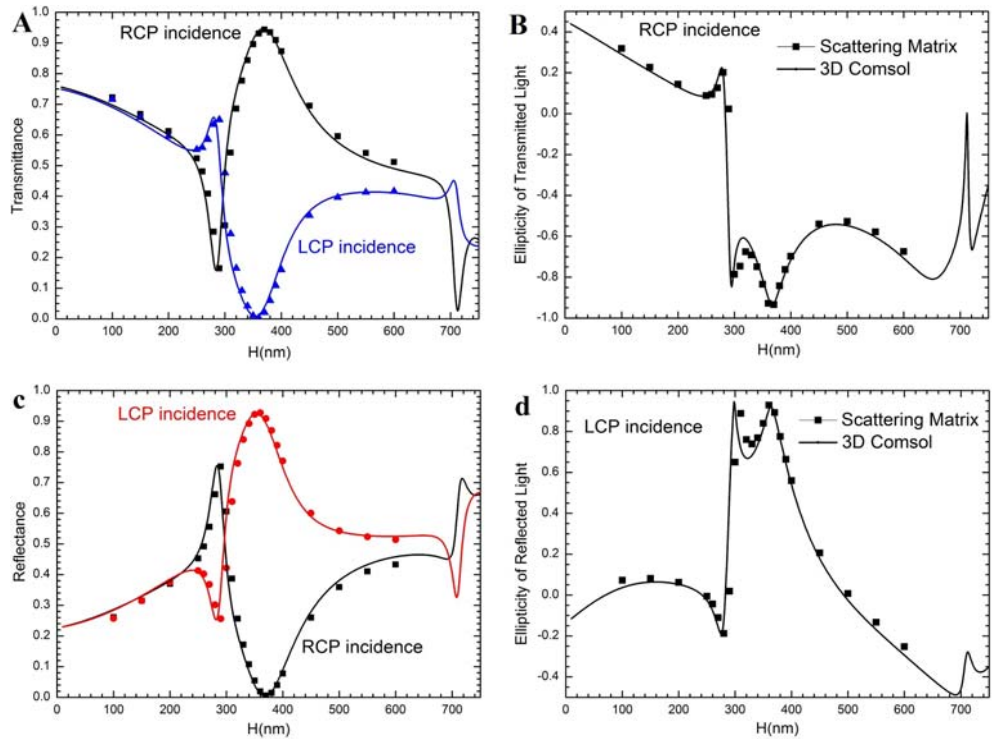

Fig.S5 Calculated spectra of circular polarizers with different thickness. Transmittance (**A**) and reflectance (**C**) of RCP and LCP with at wavelength 1500nm at normal incidence to the proposed circular polarizer. (**B**) Ellipticity of transmitted light for RCP at normal incidence. (**D**) Ellipticity of reflected light for LCP at normal incidence. The lines and symbols (such as square, trigon and circle) are results calculated by using scattering matrix and COMSOL, respectively.

4. Circular Polarizer Consisting of L-shaped Gold Particles with Trapezoid Cross Section

Considering that the cross sections of the fabricated L-shaped gold particles are trapezoid, we simulate a periodic planar array of asymmetrical L-shaped gold particles with a trapezoid cross section buried in $SiO_2$ with COMSOL. Based on the fabricated sample, geometrical parameters of trapezoid L-shaped gold particles are chosen. Figure S6A, B show calculated transmittance and relative polarization transmittance spectra of RCP and LCP at normal incidence from the top. Because the trapezoid cross section is asymmetrical, the transmittance of RCP to RCP is different with that of LCP to LCP in Fig.S6B. And, the minimum transmittances of RCP to RCP, LCP to RCP, and LCP to LCP appear at different wavelengths, which means the transmissions of them are not suppressed efficiently. So, we obtain that the planar chiral structure with trapezoid cross section can also work as a circular polarizer, but the working bandwidth is decreased significantly.

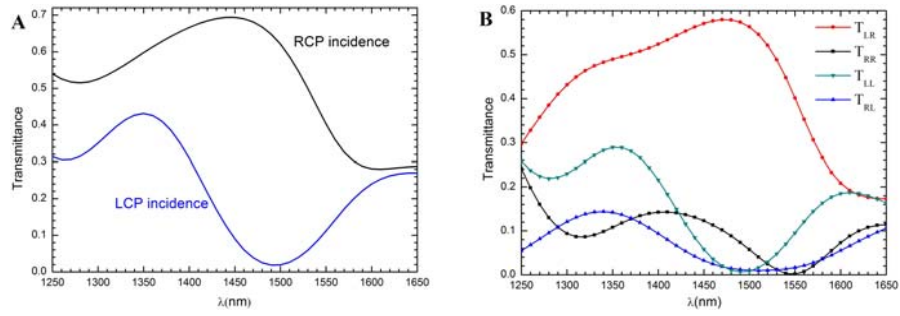

Fig.S6 Calculated spectra of a circular polarizer consisting of L-shaped gold particles with a trapezoid cross section. Transmittance (**A**) and relative polarization transmittance (**B**) spectra of RCP and LCP at normal incidence from the top to the circular polarizer consisting of a periodic planar array of asymmetrical L-shaped gold particles with the trapezoid cross section. Geometrical parameters of trapezoid L-shaped gold particles buried in $SiO_2$ are a=730nm, H=360nm, at the bottom L=580nm, Wx=246nm, Wy=160nm, at the top L=422nm, Wx=179nm, Wy=116nm, respectively.